\documentclass[acus,boxit]{JAC2001}
\usepackage{graphicx}
\begin{document}

{
\onecolumn
\thispagestyle{empty}
\renewcommand{\thefootnote}{\fnsymbol{footnote}}
\setcounter{footnote}{0}


\begin{flushright}
{\normalsize
SLAC-PUB-8875\\
Revised\\
July 2001}
\end{flushright}

\vspace{.8cm}

\begin{center}
{\bf\Large
Intrabeam Scattering Analysis of ATF Beam Measurements
}\footnote{Work supported by Department of Energy
contract DE--AC03--76SF00515.}

\vspace{1cm}

{\large
K.L.F. Bane\\
{\it Stanford Linear Accelerator Center, Stanford University,}\\
{\it Stanford, CA 94309 USA}\\\vspace{3mm}
H. Hayano, K. Kubo, T. Naito, T. Okugi, J. Urakawa\\
{\it High Energy Accelerator Research Organization (KEK),}\\
{\it 1-1 Oho, Tsukuba, Ibaraki, Japan}}

\end{center}

\vspace{.5cm}

\abstract{
At the Accelerator Test Facility (ATF) at KEK intrabeam scattering
(IBS) is a strong effect for an electron machine. It is an effect that couples
all dimensions of the beam, and in April 2000, over a short
period of time, all dimensions were measured as functions of current.
In this report we
derive a simple relation for the growth rates of emittances due
to IBS.
We apply the theories of Bjorken-Mtingwa, Piwinski, and a formula due to
Raubenheimer to the ATF parameters,
and find that the results all agree
(if in Piwinski's formalism we replace $\eta^2/\beta$
by ${\cal H}$).
Finally, we compare theory, including the effect of potential well bunch lengthening,
with the April 2000 measurements,
and find reasonably good agreement
in the energy spread and horizontal emittance dependence on current.
The vertical emittance measurement, however, implies
that either:
there is error in the
measurement (equivalent to
an introduction of 0.6\% $x$-$y$ coupling error), or
the effect of intrabeam scattering is stronger than
predicted (35\% stronger in growth rates).
}

\vfill

\begin{center}
{\it Presented at the}
{\it IEEE Particle Accelerator Conference (PAC2001),} \\
{\it Chicago, Illinois}\\
{\it June 18-22, 2001}
\end{center}

}
\setcounter{footnote}{0}

\title{INTRABEAM SCATTERING ANALYSIS OF ATF BEAM
MEASUREMENTS\vspace{-5mm}\thanks{Work supported by the Department of Energy,
contract DE-AC03-76SF00515}}
\author{K.L.F. Bane, SLAC, Stanford, CA 94309, USA\\
H. Hayano, K. Kubo, T. Naito, T. Okugi, J. Urakawa, KEK, Tsukuba,
Japan}

\maketitle

\section{INTRODUCTION}

In future e+e- linear colliders, such as the JLC/NLC, damping
rings are needed to generate beams of intense bunches with very
low emittances.  The
Accelerator Test Facility (ATF)\cite{ATF} at KEK
is a prototype for such damping rings.
In April 2000 the single bunch energy spread, bunch
length, and horizontal and vertical emittances of the
beam in the ATF were all
measured as functions of current\cite{emittanceb},\cite{ATFsigz}.
One surprising outcome was that,
at the design current, the vertical emittance appeared to
have grown by a factor of 3 over the zero-current result.
A question with important implications for the JLC/NLC is:
Is this growth real, or is it measurement error?
And if real, is it consistent with expected physical
effects, in particular,
with the theory of intrabeam scattering (IBS).

IBS is an important research topic
for many present and future low-emittance
storage rings, and
the ATF is an ideal machine for studying this topic.
In the ATF as it is now, running below design energy and with
the wigglers turned off, IBS is relatively strong for an electron machine.
It is an effect that couples all dimensions of the beam, and
at the ATF all beam dimensions can be measured.
A unique feature of the ATF is that the beam energy spread,
an especially important parameter in IBS theory,
can be measured to an accuracy of a few percent.
The bunch length measurement is important since at the ATF
potential well bunch lengthening is significant\cite{ATFsigz}.
Evidence that we are truly seeing IBS
at the ATF include (see also Ref.~\cite{Kubo}):
(1)~when moving onto the coupling resonance,
the normally large energy spread growth with current becomes
negligibly small; (2)~if we decrease the
vertical emittance using dispersion correction,
the energy spread increases.

Calculations of IBS tend to use the equations of Piwinski\cite{handbook} (P)
or of Bjorken and Mtingwa\cite{BM} (B-M).
Both approaches solve the
local, two-particle Coulomb scattering problem under certain
assumptions, but the results appear to be different.
The B-M result is thought to be the more accurate of the two, with
the difference to the P result noticeable when
applied to very low emittance
storage rings\cite{Piwref}.
Another, simpler formulation is due to Raubenheimer (R)\cite{Torth}.
Also found in the literature is a more complicated
result that allows for $x$-$y$ coupling\cite{Piw2},
and a recent formulation that includes effects of the impedance\cite{Marco}.
An optics computer program that solves IBS, using the B-M equations,
is SAD\cite{SAD}.

Calculations of IBS tend to be applied to proton or heavy ion
storage rings, where effects of IBS are normally more pronounced.
Examples of comparisons of IBS theory with measurement
can be found for proton\cite{Conte},\cite{Evans} and
electron machines\cite{CKim0},\cite{CKim}.
In such reports, although good agreement is often found,
the comparison/agreement is usually not complete ({\it e.g.} in Ref.~\cite{Conte}
growth rates agree reasonably well in the longitudinal and horizontal,
but completely disagree in the vertical)
and/or a fitting or ``fudge'' factor is
needed to get agreement
({\it e.g.} Ref.~\cite{CKim}).

In the present report we briefly describe
IBS calculations, and derive a theorem concerning the
relative vertical to horizontal IBS emittance growths
in electron machines.
We then compare the results of the P, B-M, and R methods, when
applied to the ATF parameters.
Finally, we compare the IBS growth in all beam dimensions,
including the effect of potential well bunch lengthening,
 for the B-M
calculation
and the ATF data of April 2000.

Note that this is a revised version of the original report.
After correcting for a $\sqrt{2}$ typo found in B-M, and after more carefully
considering the Coulomb log factor, the agreement
between measurement and theory has improved.

\section{IBS CALCULATIONS}

We begin by sketching the general method of calculating
the effect of IBS in a storage ring (see, {\it e.g.} Ref.~\cite{handbook}).
Let us first assume that there is no $x$-$y$ coupling.

Let us consider the IBS growth rates in energy $p$, in the
horizontal $x$, and in the vertical $y$ to be
defined as
\begin{equation}
{1\over T_p}={1\over\sigma_p}{d\sigma_p\over dt}\ ,\quad {1\over
T_x}={1\over\epsilon_x^{1/2}}{d\epsilon_x^{1/2}\over dt}\ ,\quad
{1\over T_y}={1\over\epsilon_y^{1/2}}{d\epsilon_y^{1/2}\over dt}\ .
\end{equation}
Here $\sigma_p$ is the rms (relative) energy spread,
 $\epsilon_x$ the horizontal emittance, and
$\epsilon_y$ the vertical emittance.
In general, the growth rates are given in both P and B-M theories in the form
(for details, see Refs.~\cite{handbook},\cite{BM}\footnote{
We believe that the right hand side of Eq.~4.17 in B-M
(with $\sigma_\eta$ equal to our $\sqrt{2}\sigma_p$) should be divided
by $\sqrt{2}$, in agreement with the recent derivation of Ref.~\cite{Marco}.}):
\begin{equation}
{1\over T_i}=\left< f_i\right>
\label{tinv_eq}
\end{equation}
where subscript $i$ stands for $p$, $x$, or $y$.
The functions $f_i$ are integrals that depend on
beam parameters, such as energy and phase space density, and
lattice properties, including dispersion
($y$ dispersion, though not originally in B-M,
can be added in the same manner as $x$ dispersion);
the brackets $\langle\rangle$ mean that the
quantity is averaged over the ring.


From the $1/T_i$ we obtain the steady-state properties for machines
with radiation damping:
\begin{equation}
\epsilon_x={\epsilon_{x0}\over 1-\tau_x/T_x}\ ,\
\epsilon_y={\epsilon_{y0}\over 1-\tau_y/T_y}\ ,\
\sigma^2_{p}={\sigma^2_{p0}\over 1-\tau_{p}/T_{p}}\ ,
\label{eqiterate}
\end{equation}
where subscript 0 represents the beam property due to synchrotron
radiation alone, {\it i.e.} in the absence of IBS, and the $\tau_i$
are synchrotron radiation damping times.
These are 3 coupled equations since all
3 IBS rise times depend on $\epsilon_x$, $\epsilon_y$, and
$\sigma_p$. Note that a 4th equation, the relation between
bunch length $\sigma_s$ and $\sigma_p$, is also implied; generally this is
taken to be the nominal (zero current) relation.



The best way to solve Eqs.~\ref{eqiterate} is to convert them
into 3 coupled differential equations, such as is done
in {\it e.g.} Ref.~\cite{CKim}, and solve for the asymptotic values.
For example, the equation
for $\epsilon_y$ becomes
\begin{equation}
{d\epsilon_y\over dt}= -{(\epsilon_y-\epsilon_{y0})\over \tau_y}
+{\epsilon_y\over T_y}\quad,\label{epsydif_eq}
\end{equation}
and there are corresponding
 equations for $\epsilon_x$ and $\sigma_p^2$.

Note that:

\begin{itemize}

\item For weak coupling, we add the term $-\kappa\epsilon_x$,
with $\kappa$ the coupling factor, into the
parenthesis of the $\epsilon_y$ differential equation, Eq.~\ref{epsydif_eq}.
\vspace{-2mm}

\item
A conspicuous difference between the P and B-M results is
their dependence on dispersion $\eta$: for P the $f_i$
depend on it only through $\eta^2$; for B-M,
through $[\eta^\prime+\beta^\prime\eta/(2\beta)]$ and
the dispersion invariant
${\cal H}=\bar{\gamma}\eta^2+2\alpha\eta\eta^\prime+\beta{\eta^\prime}^2$,
with $\alpha$, $\beta$, $\bar\gamma$ Twiss parameters.
\vspace{-2mm}

\item At the ATF, at the
highest single bunch currents,
there is significant potential well bunch lengthening,
though we are still below the threshold to the microwave instability\cite{ATFsigz}.
We can approximate
the bunch lengthening effect in our IBS calculations by adding a
multiplicative factor $f_{pw}(I)$
[$I$ is current], obtained from
measurements,
to the equation relating $\sigma_s$ to $\sigma_p$.
\vspace{-2mm}

\item The results include a so-called Coulomb log factor, of the form
$\ln(b_{max}/b_{min})$, where $b_{max}$, $b_{min}$ are the
maximum, minimum impact parameters, quantities which are not well
defined; typically $\ln()\sim20$.
For typical, flat beams we take
$b_{max}$ to be the vertical beam size, $\sigma_y$;
$b_{min}=r_0 c^2/\langle v_x^2\rangle=r_0\beta_x/(\gamma^2\epsilon_x)$,
with $r_0$ the classical
electron radius ($2.82\times10^{-15}$~m), $c$ the speed of light,
$v_x$ the transverse velocity in the rest frame, and
$\gamma$ the energy factor.
For the ATF, $\ln()=16.0$.
\vspace{-2mm}

\item The IBS bunch distributions
are not Gaussian, and tail particles can be overemphasized in these
solutions.
We are interested in core sizes, which we estimate
by eliminating interactions with collision rates
less than the synchrotron radiation damping rate\cite{Tor}.
We can approximate this in the Coulomb log term by
letting
$\pi b_{min}^2\langle |v_x|\rangle\langle n\rangle=1/\tau$,
with
$n$ the particle density in the rest frame\cite{SAD2};
or
$b_{min}=\sqrt{4\pi\sigma_x\sigma_y\sigma_z/[Nc\tau]}
(\beta_x/\epsilon_x)^{1/4}$, with
$N$ the bunch population.
For the ATF
with this cut, $\ln()=13.9$.
\vspace{-2mm}

\end{itemize}

\subsection{Emittance Growth}

An approximation to Eqs.~\ref{tinv_eq}, valid for
typical, flat electron beams is
due to Raubenheimer (R)
\cite{Torth},\cite{Piw_ana}:
\footnote{Our equation for $1/T_p$
is twice as large
as Eq.~2.3.5 of Ref.~\cite{Torth}.}
\begin{eqnarray}
{1\over T_p}& \approx &
{r_0^2 cN\over 32\gamma^3\epsilon_x\epsilon_y\sigma_s\sigma_p^2}
\left({\epsilon_x\epsilon_y\over\langle\beta_x\rangle\langle\beta_y\rangle}\right)^{1/4}
\ln\left({\langle\sigma_y\rangle\gamma^2\epsilon_x\over r_0\langle\beta_x\rangle}\right)\nonumber\\
{1\over T_{x,y}}& \approx &
{\sigma_p^2\langle{\cal H}_{x,y}
\rangle\over\epsilon_{x,y}}{1\over T_p}\quad.\label{Tor_eq}
\end{eqnarray}
If the vertical emittance is due only to vertical dispersion
then\cite{Torth}
\begin{equation}
\epsilon_{y0}\approx {\cal J}_\epsilon\langle{\cal H}_y\rangle\sigma_{p0}^2\quad,
\label{Tor0_eq}
\end{equation}
with ${\cal J}_\epsilon$ the energy damping partition number.
We can solve Eqs.~\ref{eqiterate},\ref{Tor_eq},\ref{Tor0_eq} to
obtain the steady-state beam sizes. Note that once the vertical
orbit---and therefore $\langle{\cal H}_y\rangle$---is set, $\epsilon_{y0}$
is also determined.

Following an argument in Ref.~\cite{Torth} we can obtain
a relation between the expected vertical and horizontal emittance
growth due to IBS in the presence of random vertical dispersion:
The beam momentum in the longitudinal plane is much less
than in the transverse planes. Therefore, IBS will first heat
the longitudinal plane; this, in turn, increases the transverse emittances
through dispersion (through ${\cal H}$),
like synchrotron radiation (SR) does.
One difference between IBS and SR is that IBS increases the emittance
everywhere, and SR only in bends. We can write
\begin{equation}
{\epsilon_{y0}\over\epsilon_{x0}}\approx
{{\cal J}_x\langle{\cal H}_y\rangle_{b}
\over{\cal J}_y\langle{\cal H}_x\rangle_{b}}
\quad ,\quad
{\epsilon_{y}-\epsilon_{y0}\over\epsilon_{x}-\epsilon_{x0}}\approx
{{\cal J}_x\langle{\cal H}_y\rangle\over{\cal J}_y\langle{\cal H}_x\rangle}\ ,
\end{equation}
where ${\cal J}_{x,y}$ are damping partition numbers, and
$\langle \rangle_{b}$ means averaging is only done over the bends.
For vertical dispersion due to errors we expect
$\langle{\cal H}_y\rangle_{b}\approx\langle{\cal H}_y\rangle$.
Therefore,
\begin{equation}
r_\epsilon\equiv
{(\epsilon_{y}-\epsilon_{y0})/\epsilon_{y0}
\over(\epsilon_{x}-\epsilon_{x0})/\epsilon_{x0}}\approx
{\langle{\cal H}_x\rangle_{b}\over\langle{\cal H}_x\rangle}
\quad,\label{Kubo_eq}
\end{equation}
which, for the ATF is 1.6.
If, however, there is only $x$-$y$ coupling, $r_\epsilon=1$;
if there is both vertical dispersion and coupling, $r_\epsilon$ will
be between $\langle{\cal H}_x\rangle_{b}/\langle{\cal H}_x\rangle$ and 1.
\vspace{-2.0mm}

\subsection{Numerical Comparison}

Let us compare the results of the P, B-M, and R
methods when applied to the ATF beam parameters and lattice,
with vertical dispersion and no $x$-$y$ coupling.
We take: current $I=3.1$~mA, energy $E=1.28$~GeV,
$\sigma_{p0}=5.44\times10^{-4}$,
$\sigma_{s0}=5.06$~mm (for an rf voltage of 300~kV),
$\epsilon_{x0}=1.05$~nm, $\tau_p=20.9$~ms, $\tau_x=18.2$~ms, and
$\tau_y=29.2$~ms; $f_{pw}=1$. The ATF circumference is 138~m,
${\cal J}_\epsilon=1.4$,
$\langle\beta_x\rangle=3.9$~m, $\langle\beta_y\rangle=4.5$~m,
$\langle \eta_x\rangle=5.2$~cm and $\langle{\cal H}_x\rangle=2.9$~mm.
To generate vertical dispersion we randomly offset magnets
by 15~$\mu$m,
and then calculate the closed orbit using SAD.
For our seed we find that the rms dispersion
$(\eta_y)_{rms}=7.4$~mm, $\langle{\cal H}_y\rangle=17$~$\mu$m,
and $\epsilon_{y0}=6.9$~pm (in agreement with Eq.~\ref{Tor0_eq}).
For consistency between the methods we here take
$\ln()=\ln{[\langle\sigma_y\rangle\gamma^2\epsilon_x/(r_0\langle\beta_x\rangle)]}=16$.

\begin{figure}[htb]
\centering
\includegraphics*[width=74mm]{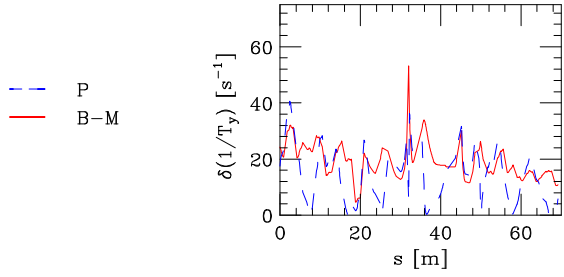}
\caption{ Vertical differential growth rate over 1/2 the ATF, for
 Piwinski (dashes) and  Bjorken-Mtingwa (solid)\vspace{-1mm}.
 }
\label{fidebug0}
\end{figure}

\vspace{-3mm}Performing the calculations,
 we find that the growth rates in $p$ and $x$ agree well between the two methods;
the vertical rate, however, does not.
Fig.~\ref{fidebug0} displays the vertical {\it differential} IBS growth
rate $\delta(1/T_y)$,
over half the ring (the periodicity is 2), as
obtained by the two methods (dashes for P, solid for B-M).
The IBS growth rate $1/T_y$ is the
average value of this function.
We see that the P curve is enveloped by the B-M curve;
the average P result is 25\% less than that of B-M.

\begin{figure}[htb]
\centering
\includegraphics*[width=82mm]{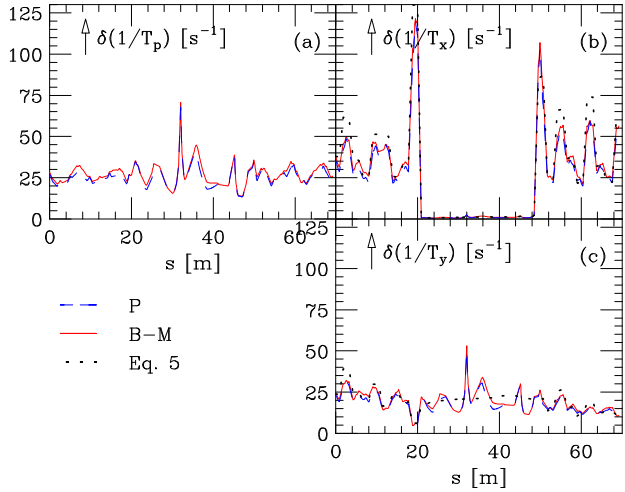}
\caption{ Differential growth rates over 1/2 the ATF, as obtained
by modified Piwinski (dashes) and  Bjorken-Mtingwa (solid).
 }
\label{fidebug}
\end{figure}

From the arguments of Sec.~2.1, we might expect that we can
improve the P calculation if we replace $\eta_{x,y}^2/\beta_{x,y}$
in the formulation by
${\cal H}_{x,y}$. Doing this we find that, indeed, the three differential growth
rates now agree reasonably well with the B-M results (see Fig.~\ref{fidebug}).
As for the averages, the P results are all systematically low, by 6\%.
According to the B-M method
$1/T_p=27.0$~s$^{-1}$, $1/T_x=26.0$~s$^{-1}$,
$1/T_y=19.4$~s$^{-1}$;  $\sigma_p/\sigma_{p0}=1.52$,
$\epsilon_x/\epsilon_{x0}=1.90$, $\epsilon_y/\epsilon_{y0}=2.30$.
The emittance ratio of Eq.~\ref{Kubo_eq} is $r_\epsilon=1.44$,
close to the expected 1.6.

The dots in Fig.~\ref{fidebug}b,c give
the differential rates corresponding to Eq.~\ref{Tor_eq},
and we see that the agreement also is good.
The growth rates in ($p$,$x$,$y$) are (27.0,26.4,19.3)~s$^{-1}$,
the relative growths in ($\sigma_p$,$\epsilon_x$,$\epsilon_y$)
are (1.51,1.92,2.29).
\vspace{-1mm}

\section{COMPARISON WITH MEASUREMENT}

\begin{figure}[htb]
\centering
\includegraphics*[width=82mm]{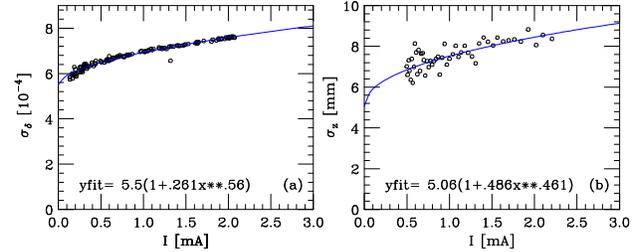}
\caption{ Measurements of energy spread~(a) and bunch length~(b),
 with $V_c=300$~kV.
 } \label{fisigpsigz}
\end{figure}

\begin{figure}[htb]
\centering
\includegraphics*[width=82mm]{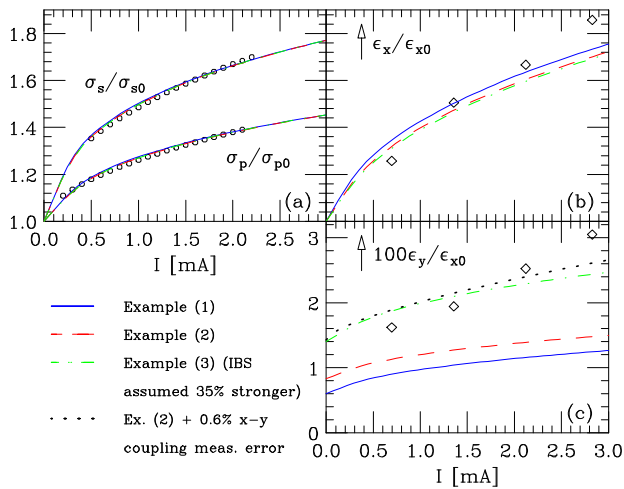}
\caption{ATF measurement data (symbols) and IBS theory fits (the
curves). The symbols in (a) give the smooth curve fits to the
measured data of Fig.~\ref{fisigpsigz}.
 } \label{fifit}
\end{figure}

The parameters $\sigma_p$, $\sigma_s$, $\epsilon_x$, and
$\epsilon_y$ were measured in the ATF as functions of current over
a short period of time at rf voltage $V_c=300$~kV.
Energy spread was
measured on a screen at
a dispersive region in the extraction line
(Fig.~\ref{fisigpsigz}a);
bunch length with
a streak camera in the ring
(Fig.~\ref{fisigpsigz}b). The curves in the plots are
fits that give the expected zero current
result.
Emittances were measured on wire monitors in the extraction
line (the symbols in Fig.~\ref{fifit}b-c; note that
the symbols in Fig.~\ref{fifit}a reproduce the fits
to the data of Fig.~\ref{fisigpsigz}).
Unfortunately, we do not have error bars for this data.
In $x$, nevertheless, we expect the errors to be small.
In $y$, from experience, we expect the random component
of errors to be 5--10\%.
As for the systematic component, it is conceivable that it is not small,
since $\epsilon_y$ is small
and it only takes a small amount of roll or dispersion in the
extraction line to significantly affect the measurement result.

We see that $\epsilon_x$
appears to grow by $\sim85\%$ by $I=3$~mA; $\epsilon_y$
begins at about 1.0-1.2\% of $\epsilon_{x0}$, and then grows to
about 3\% of $\epsilon_{x0}$, implying that $r_\epsilon=1.8$--2.4.
If we are
vertical dispersion dominated, with
$(\eta_y)_{rms}=10$~mm and $\epsilon_{y0}\approx.012\epsilon_{x0}$,
then the data nearly satisfies
Eq.~\ref{Kubo_eq}, $r_\epsilon\approx1.6$.
However, normally, after dispersion correction, the residual
dispersion at the ATF is kept to $(\eta_y)_{rms}=3$--5~mm.
On the other hand, if
 we are coupling dominated we see that $r_\epsilon\approx1$ is not well satisfied
by the data.

Let us compare B-M calculations with the data.
Here we take $f_{pw}$ as given by the measurements, and
take $\ln()=14$. At $I=3$~mA we adjust $\epsilon_{y0}$
until the calculated $\sigma_p$ agrees with the measurement.
In Fig.~\ref{fifit} we give examples:
(1)~with vertical dispersion only, with $(\eta_y)_{rms}=7.0$~mm
 and $\epsilon_{y0}=6.3$~pm
(solid);
(2)~coupling dominated with $(\eta_y)_{rms}=3$~mm
and $\epsilon_{y0}=8.7$~pm
(dashes);
(3)~increasing the strength of IBS by increasing $\ln()$
by 35\%: {\it i.e.} letting
$\ln()=19$, for the coupling dominated example with $(\eta_y)_{rms}=3$~mm
and $\epsilon_{y0}=14.7$~pm
(dotdash);
(4)~same as Ex.~2 but assuming a small amount of
$\epsilon_y$ measurement error, {\it i.e.} adding 0.6\% $x$-$y$ coupling error (the dots).

We see that,  for all examples,
$\sigma_p(I)$ agrees well with measurement, and $\epsilon_x(I)$
agrees reasonably well also.
For general agreement for $\epsilon_y(I)$, we need either a small
amount of measurement error
({\it e.g.} 0.6\% $x$-$y$ coupling measurement error),
 or for IBS to be 35\% stronger than expected.
A difference in $\ln()$ of 5 units implies a factor of 150 in the argument.
Although there is uncertainty in the Coulomb log factor, this difference
seems larger than we expect the uncertainty to be.
Note that the expected error in the IBS calculation itself, assuming
$\ln()$ is correct, is also small: $\sim1/\ln()=5\%$\cite{Marco}.
And finally, note that even if we can account for the offset by {\it e.g.}
a 0.6\% $x$-$y$ coupling measurement error,
we see from Fig.~\ref{fifit}
that the slope of the vertical emittance dependence on current is still steeper
than predicted.

\section{CONCLUSION}

We have derived a simple relation for relative growth rates
of emittances due to IBS.
We have found that for the ATF, IBS calculations following
Piwinski (with $\eta^2/\beta$ replaced by ${\cal H}$),
Bjorken-Mtingwa, and a formula due to
Raubenheimer
all agree well (though one needs to
be consistent in choice of Coulomb log factor).

Comparing the Bjorken-Mtingwa
calculations (including the effect of potential well bunch lengthening)
with the ATF measurements of April 2000,
we have found reasonably good agreement
in the energy spread and horizontal emittance dependence on current.
The vertical emittance measurement, however, implies
that either:
there is error in the
measurement (equivalent to
an introduction of 0.6\% $x$-$y$ coupling error), or
the effect of intrabeam scattering is stronger than
predicted (35\% stronger in growth rates).
In addition, the slope of the vertical emittance dependence on
current is steeper than predicted.


We thank A. Piwinski for help in understanding IBS
and K. Oide for explaining IBS calculations in SAD.\vspace{-1mm}
\vspace{-2.0mm}

\pagebreak

\end{document}